\documentclass[multphys,vecphys,url]{svmult}

\usepackage{makeidx}         
\usepackage{graphicx}        
\usepackage{multicol}        
\usepackage[bottom]{footmisc}

\makeindex             

\begin{document}
\title*{Anomalous thermal escape in Josephson systems perturbed by microwaves}
\titlerunning{Anomalous thermal escape in Josephson systems.....}
\author{N.~Gr{\o }nbech-Jensen\inst{1} \and
M. G. Castellano\inst{2} \and
F. Chiarello\inst{2} \and
M. Cirillo\inst{3} \and
C. Cosmelli\inst{4} \and
V. Merlo\inst{3} \and
R. Russo\inst{3,5} \and
G. Torrioli\inst{2}}
\institute{Department of Applied Science, University of California, Davis, California 95616, U.S.A. \and
IFN-CNR and INFN, via Cineto Romano 42, I-00156 Rome, Italy. \and
Department of Physics and INFM, University of Rome "Tor Vergata", I-00133 Rome, Italy. \and
Department of Physics and INFN, University of Rome, "La Sapienza", I-00185 Rome, Italy.
\and Istituto di Cibernetica "E.Caianiello", CNR, I-80078 Pozzuoli, Italy.
}
\authorrunning{N.~Gr{\o }nbech-Jensen {\it et al.}}
%
%
\maketitle

{\small{\it
We investigate, by experiments and numerical simulations, thermal activation
processes of Josephson tunnel junctions in the presence of microwave
radiation. When the applied signal resonates with the Josephson plasma
frequency oscillations, the switching current may become multi-valued
temperature ranges both below and above the the classical to quantum crossover
temperature. Switching current distributions are obtained both experimentally
and numerically at temperatures both near and far above the quantum crossover
temperature. Plots of the switching currents traced as a function of the
applied signal frequency show very good agreement with a simple anharmonic
theory for Josephson resonance frequency as a function of bias current.
Throughout, experimental results and direct numerical simulations of
the corresponding thermally driven classical Josephson junction model
show very good agreement.}}

\normalsize
\section{Introduction}
\label{sec:1}
The Josephson tunnel junction is a physical system very well
studied due to its simplicity and nonlinearity \cite{Barone82}.
Statistical properties of Josephson junctions have been another
subject of intense investigation through, e.g., measurements of the escape
statistics from the zero-voltage state,
successfully confirming consistency with the classic Kramers model for
thermal activation from a potential well \cite{Kramers40,Kramers_stuff}.
Escape measurements represent a powerful tool for probing the nature
of the underlying potential well, and
applying an ac field to a low-temperature system has been reported to
produce anomalous switching distributions with two, or more,
distinct dc bias currents for which switching is likely. These measurements
have been interpreted as a signature of the ac field aiding the population
of multiple quantum levels in a junction, thereby leading to enhancement of
the switching probability for bias currents for which the corresponding
quantum levels match the energy of the microwave photons. Work performed in
this direction appeared first in the literature two decades ago\cite
{Martinis1,Martinis2}. These results have significantly attracted
interest toward Josephson junction systems as possible basic elements in the
field of quantum coherence and quantum computing\cite
{Ruggiero99,Martinis3,Han02,Berkley03,Ustinov03}, and more recently
other investigations have further indicated that the application of microwaves
may not be the only condition under which level
quantization can be observed in Josephson junctions \cite{Silvestrini97}.

Within the framework of
this research topic we recently reported experimental measurements
conducted on a Josephson junction, operated
well above the so-called quantum transition
temperature $T_*$, and direct numerical simulations of the classical
pendulum model, parameterized to mimic the experimental device\cite{Jensen_04}.
It was found that multi-peaked switching distributions are not unique to
the quantum regime (below $T_*=\hbar\omega_0/2\pi k_B$), and, in fact,
are manifested with the same features and under the same conditions 
in the classical regime as has been previously reported for low temperature
measurements below $T_*$.
With the present paper we wish to contribute an anharmonic theory that
accurately captures the bias current values of the observed resonant
peaks in the switching distributions as a function of the applied frequency
of the microwave field. We demonstrate agreement between the theory,
direct numerical simulations, and experimental measurements for 
direct resonances as well as harmonic and sub-harmonic resonances at
temperatures both well above and near $T_*$.

\begin{figure}
\centering
\includegraphics[width=3.5in,angle=0]{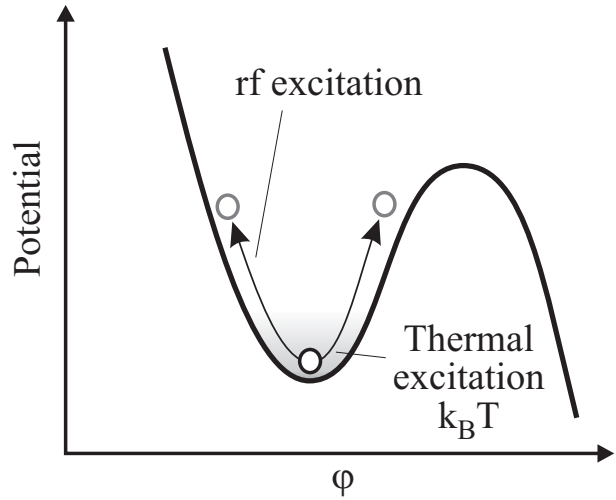}\hspace*{0.5cm}

%
%
\caption{Sketch of the physical phenomenon under investigation: a driven
oscillation energy $E_{ac}$ superimposed onto thermal excitations, may cause
a particle to escape a washboard potential.}
\label{fig:1}       
\end{figure}

Figure 1 illustrates the process under investigation: in the classical
one-degree-of-freedom single-particle washboard potential of the Josephson
junction \cite{Martinis1}, thermal excitations (shaded in the sketch) of
energy $k_BT$ and the energy $E_{ac}$ of forced oscillations due to
microwave radiation, can cause the particle to escape from the potential
well. This process can be traced by sweeping the current-voltage
characteristics of the Josephson junction periodically. Escape from the
potential well corresponds to an abrupt transition from the top of the
Josephson-current zero-voltage state to a non-zero voltage state. The
statistics of the switching events, in the absence of time-varying
perturbations, have been shown to be consistent with Kramers' model \cite
{Kramers40} for thermal escape from a one-dimensional potential. Since the
thermal equilibrium Kramers model does not include the effect of
non-equilibrium force terms, the results of the switching events generated
by the presence of a microwave radiation on a Josephson junction can be
investigated, in a thermal regime, only by a direct numerical simulation of
the governing equations (RSCJ model) \cite{Barone82}.
\section{Theory}
The RSCJ model reads,
\begin{equation}
\frac{\hbar C}{2e}\frac{d^2\varphi }{dt^2}+\frac \hbar {2eR}\frac{d\varphi }{%
dt}+I_c\sin \varphi =I_{dc}+I_{ac}\sin \omega _dt+N(t)\;.  \label{eq:Eq_1}
\end{equation}
Here, $\varphi $ is the phase difference of the quantum mechanical wave
functions of the superconductors defining the Josephson junction, $C$ is the
magnitude of junction capacitance, $R$ is the model shunting resistance, and 
$I_c$ is the critical current, while $I_{dc}$ and $I_{ac}\sin \omega _dt$
represent, respectively, the continuous and alternating bias current flowing
through the junction. 
The term $N(t)$ represents the thermal noise-current due
to the resistor $R$ given by the thermodynamic dissipation-fluctuation
relationship \cite{Parisi88} 
\begin{eqnarray}
\left\langle N(t)\right\rangle &=&0  \label{eq:Eq_2} \\
\left\langle N(t)N(t^{\prime })\right\rangle &=&2\frac{k_BT}R\delta
(t-t^{\prime })\;,  \label{eq:Eq_3}
\end{eqnarray}
with $T$ being the temperature. The symbol, $\delta (t-t^{\prime })$, is the
Dirac delta function. Current and time are usually normalized respectively
to the Josephson critical current $I_c$ and to $\omega _0^{-1}$, where $%
\omega _0=\sqrt{2eI_c/\hbar C}$ is the Josephson plasma frequency. With this
normalization, the coefficient of the first-order phase derivative becomes
the normalized dissipation $\alpha =\hbar \omega _0/2eRI_c$. It is also
convenient to scale the energies to the Josephson energy $E_J=I_c\hbar
/2e=I_c\Phi _0/2\pi $, where $\Phi _0=h/2e=2.07\cdot 10^{-15}Wb$ is the
flux-quantum. Thus, the set of equations (1-3) can be expressed in
normalized form as 
\begin{eqnarray}
\ddot \varphi +\alpha \dot \varphi +\sin \varphi &=&\eta +\eta _d\sin \Omega
_d\tau +n(\tau )  \label{eq:Eq_4} \\
\left\langle n(\tau )\right\rangle &=&0  \label{eq:Eq_5} \\
\left\langle n(\tau )n(\tau ^{\prime })\right\rangle &=&2\alpha \theta
\delta (\tau -\tau ^{\prime })\ \;,  \label{eq:Eq_6}
\end{eqnarray}
where $\theta =\frac{k_BT}{E_J}$ is the normalized temperature.
The normalized dc and ac currents are
$\eta = \frac{I_{dc}}{I_c}$ and $\eta _d=\frac{I_{ac}}{I_c}$, respectively.

For small-amplitude oscillations around a stable (zero-voltage)
energetic minimum we obtain the standard relationship between
resonance frequency and bias current,
\begin{eqnarray}
\Omega_p & = & \sqrt[4]{1-\eta^2} \; , \label{eq:Eq_wh}
\end{eqnarray}
where we have omitted the dissipative contribution to the resonance
frequency. However, this linear resonance is not directly relevant
for the dynamics leading to anomalous resonant switching. Looking
at Figure 1 it is obvious that a switching event will arise from
probing the anharmonic region of the potential near the local energetic
maximum, and we therefore must anticipate a depression of the resonance
frequency at these large amplitudes. In order to quantify this notion,
we will adopt the following ansatz,
\begin{eqnarray}
\varphi & = & \varphi_0+\psi \; , \label{eq:Eq_ansatz}
\end{eqnarray}
where $\varphi_0$ is a constant and $\psi$ represents oscillatory motion.
Inserting this ansatz into equation (\ref{eq:Eq_4}) (for $\theta=0$)
yields,
\begin{eqnarray}
\ddot{\psi}+\sin\varphi_0\cos\psi+\cos\varphi_0\sin\psi & = & \eta+\eta_d\sin\Omega_d 
t-\alpha\dot{\psi} \; . \label{eq:Eq_nonl}
\end{eqnarray}
Making the single-mode assumption, $\psi=a\sin(\Omega_d t+\kappa)$,
$\kappa$ being some constant phase,
we obtain the following
\begin{eqnarray}
\sin\varphi_0 & = & \frac{\eta}{J_0(a)} \label{eq:Eq_static} \\
\Omega_{res} & = & \sqrt{\frac{2J_1(a)}{a}\sqrt{1-\left(\frac{\eta}{J_0(a)}\right)^2}} \; , \label{eq:Eq_anharm}
\end{eqnarray}
the functions, $J_n$, being the zero's order Bessel function of the first kind.
Notice that $\Omega_{res}\rightarrow\Omega_p$ for $a\rightarrow0$.
Since multi-peaked switching distributions must require some switching events
to happen near the resonance and others to happen for larger bias currents,
we can estimate that the amplitude $a$ must be approximately given by
\begin{eqnarray}
a & \approx & \pi-2\sin^{-1}\frac{\eta}{J_0(a)}
\end{eqnarray}
which represents the phase distance from the energetic minimum to the saddle
point. Approximating $J_0(a)$ by its Taylor expansion, we can arrive at
the simple expression between the oscillation amplitude and the applied 
bias current $\eta$,
\begin{eqnarray}
a & \approx & \sqrt{\frac{4}{3}\left[1-\eta\right]} \; .
\end{eqnarray}
Inserting this approximate expression into (11) gives an explicit
relationship between the anharmonic resonance and the bias current,
relevant for the bias current location of the anomalous secondary
peak in the switching distribution.

\section{Experiments and Simulations}
Experiments were performed on Josephson tunnel junctions
fabricated according to classical Nb-NbAlOx-Nb procedures \cite{fabrication}%
. The samples had very good current-voltage characteristics and magnetic
field diffraction patterns. The junctions were cooled in a $^3$He
refrigerator (Oxford Instruments Heliox system), providing temperatures down
to 360mK. Microwave radiation, brought to the chip-holders by a coax
cable, was coupled capacitively to the junctions, and the junction had a maximum
critical current of $I_c=143\mu A$ and a total capacitance of $6pF$ from which
we estimate a plasma frequency of $\omega_0/2\pi=42.5GHz$. From this
value of the
plasma frequency the classical to quantum crossover temperature 
\cite{Affleck} $T_{*}=(\hbar \omega _0/2\pi k_B)=325mK$ between classical
thermal and quantum mechanical behavior can be estimated. The sweep rate of
the continuous current $I_{dc}$ was $\dot I_{dc}=800mA/s$, and we verified
that the experiment was being conducted in adiabatic conditions \cite
{Silvestrini97}. The junction has a Josephson energy
$E_J\approx46.4\cdot 10^{-21}J$ in the temperature range from 370mK to 1.6K,
and effective resistance $R=74\Omega$.
Evaluation of the dissipation parameter was based on the
hysteresis of the current-voltage characteristics of the junctions\cite
{Barone82}.
We show data for two temperatures, $T\approx 388mK$ and $T=1.6K$.

Figure 2 shows experimentally obtained results at $T=1.6K$ \cite{Jensen_04}.
The lower frames of the figure displays the switching distributions
in bias current at different microwave frequencies. The top frame shows
the relationship between the normalized current, for which the switching
distributions
have their resonant peak, and the applied frequency (normalized to the
junction plasma frequency). Each black marker represents one of the 
switching distributions. Also shown in figure 2 is the linear resonance of
equation (7), sown as a dashed curve, and the anharmonic resonance of
equations (11) and (13), shown as a solid curve. The agreement between
the experimental measurements and the anharmonic theory of the classical
model is near perfect for the available data points, and we emphasize that
the theoretical model of equations (11) and (13) has no fitted parameters
to adjust in the comparison. Thus, the consistent depression of the 
experimental data relative to a linear resonance consideration, observed
in Ref.~\cite{Jensen_04}, should be expected and not give rise to
re-fitting the critical current or the plasma resonance frequency.

\begin{figure}
\centering
\includegraphics[width=3.5in,angle=90]{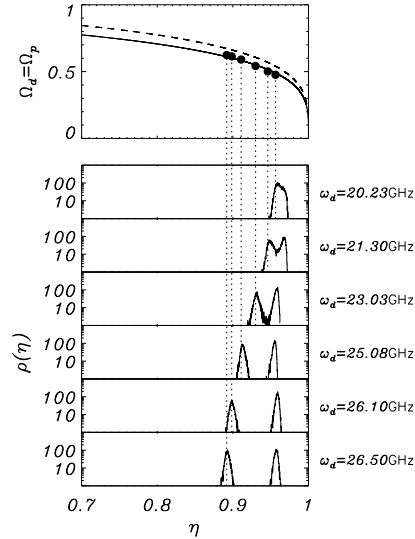}\hspace*{0.5cm}
%
%
\caption{Experimentally obtained switching distributions, $\rho(\eta)$,
for the microwave-driven
junction obtained for increasing values of the drive frequency. The
frequency data points in the uppermost plot are relative to the position of
the secondary peak in the plots. Temperature is $T=1.6K$, and bias sweep rate
is $\dot{I}=800mA/s$.
Dashed curve in uppermost graph
represents the linear plasma resonance of (7), while the solid curve represents
the anharmonic resonance of (11) and (13).}
\label{fig:2}       
\end{figure}

Figure 3 shows experimentally obtained results at $T=388mK$, presented
in the same manner as the data in figure 2. The agreement between the
experimental measurements and the anharmonic theory of the classical
model is again near perfect for the available data points. The resonance
curves shown in figures 2 and 3 are identical, since we have not included
the resonance dependence on the dissipation (since dissipation is very small)
and since the measurements indicated that plasma resonance frequency and 
critical current were unchanged in the investigated temperature range.
By comparing figures 2 and 3, we notice that there seem to be no qualitative
(and hardly any quantitative) differences between the data obtained at the
two very different temperatures, even though the data of figure 2 is
acquired at $T\approx5T_*$ and the data in figure 3 represent $T\approx1.2T_*$.

\begin{figure}
\centering
\includegraphics[width=3.5in,angle=90]{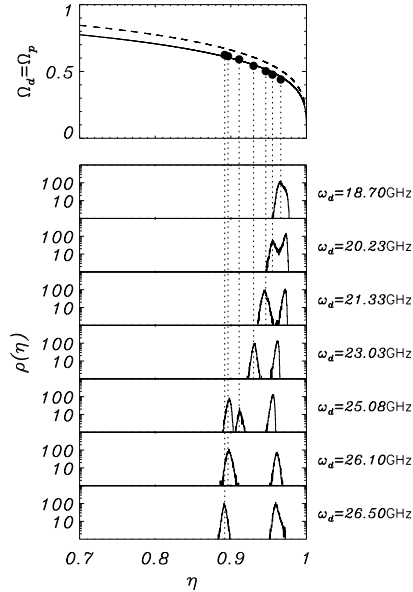}\hspace*{0.5cm}
%
%
\caption{Same experimental situation as described in figure 2, but
data is here acquired at $T=388mK$.}
\label{fig:3}       
\end{figure}

Numerical simulations of escape in a system described by equations (4)-(6)
corresponding to the experiments
with $\alpha =0.00845$, $\theta =115.4\cdot10^{-6}, 4.76\cdot 10^{-4}$,
and continuous bias
sweep rate $\frac{d\eta }{d\tau }=2.1\cdot 10^{-8}$ have also been
conducted in order to investigate the purely classical dynamics in
comparison with the experimental measurements. The parameters have
been chosen in agreement with the experiments discussed above. Switching
distributions (each corresponding to 1,000-10,000 events), obtained
for different values of the normalized drive frequency and temperature,
were obtained as a function of the continuous bias, and secondary 
resonant peaks in the distribution were easily obtained in the
classical model by adjusting the simulated microwave amplitude for
a given frequency.

\begin{figure}
\centering
\includegraphics[width=2.4in,angle=90]{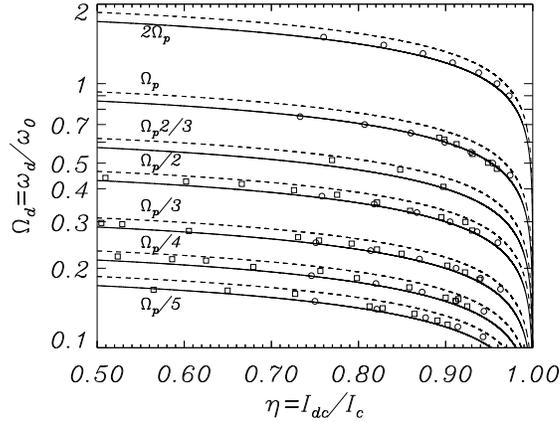}\hspace*{0.5cm}
%
%
\caption{The functional dependencies of the driving frequency upon the
location of the secondary peak in $\rho(\eta)$ obtained for subharmonic and
harmonic pumping. Circles represent numerical results and squares
experimental data. Parameters are as given in Figure 2.
Dashed curvesrepresent the linear plasma resonance of (7), while the solid
curve s represents the anharmonic resonance of (11) and (13).}
\label{fig:4}       
\end{figure}

Figure 4 shows both experimental measurements and direct numerical
simulations of the resonant peak location as a function of applied
microwave frequency at $T=1.6K\approx5T_*$.
Experimental data are shown as box markers
and numerically obtained data are shown as circles. As in figures
2 and 3, dashed curves represent the linear resonance (7) while the solid
curves are generated from (11) and (13). Experimental data for the fundamental
resonance (labeled $\Omega_p$) in the figure are the ones from figure 2.
We clearly observe the close agreement between theory, experiment,
and simulation. We also present data for subharmonic resonances, and
here too do we find very close agreement between simulation and experiment. The
theoretical harmonic and subharmonic resonance curves are the ones of equations
(7), (11), and (13) multiplied with the indicated fraction in the figure.
This simple theory seems to also predict the sub-harmonic microwave
induced resonant peak location very well.

\begin{figure}
\centering
\includegraphics[width=2.4in,angle=90]{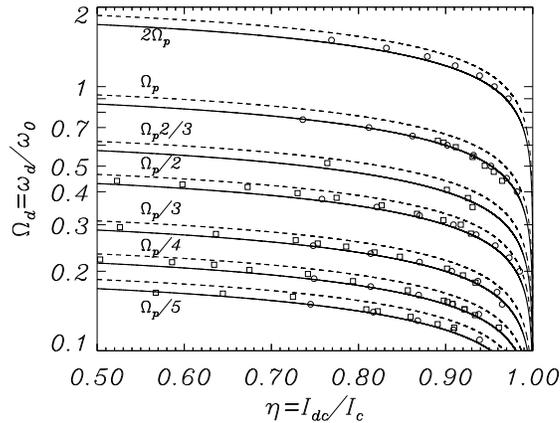}\hspace*{0.5cm}
%
%
\caption{The functional dependencies of the driving frequency upon the
location of the secondary peak as described in figure 4. The temperature is
here $T=388mK$.}
\label{fig:5}       
\end{figure}
We finally show, in figure 5, the data similar to the ones in figure 4
taken at $T=388mK\approx1.2T_*$. Also at this temperature do we observe very
close agreement between experiment, simulation, and theory, amplifying the
notion that microwave induced switching and anomalous switching
distributions can be understood within 
a classical, thermal framework.

\section{Conclusion}
In conclusion, our theory, and experiments on ac-driven, thermal
escape of a {\it %
classical} particle from a one-dimensional potential  well have shown that
resonant coupling (harmonic or subharmonic) between the applied microwaves
and the plasma resonance frequency provides an enhanced opportunity for
escape, and we have directly observed the signatures of such
microwave-induced escape distributions in the form of anomalous multi-peaked
escape statistics at two temperatures, $T=388mK\approx1.2T_*$ and
$T=1.6K\approx5T_*$. The straightforward agreement between the classical
hypothesis of anomalous distributions being directly produced by ac-induced
anharmonic resonances, the results of numerical simulations of the
classical pendulum
model of a Josephson junction, and actual Josephson junction experiments
indicate a consistent interpretation of ac-induced anomalous multi-peaked
switching distributions in the classical regime of Josephson junctions.

It is noted that previous experimental work on ac-induced escape
distributions obtained at temperatures below $T_{*}$ is consistent with the
observations presented here. Those experiments have produced ac-induced
peaks in the observed switching distributions, and the relevant peaks are
located alongside the expected classical plasma resonance curve, as we have
also found here. An important observation is that the microwave-radiation
frequency necessary for populating an excited quantum level ($\hbar \omega
_d $) in a quantum oscillator coincides with the classical resonance
frequency of the corresponding classical oscillator. Thus, the switching
distributions obtained from classical and quantum mechanical oscillators may
exhibit the same microwave induced multi-peak signatures, which in the
classical interpretation is merely due to resonant nonlinear effects. It is
evident then that multi-peaked switching distributions
are not a unique signature of quantum
behavior in the ac-driven Josephson junction. We finally point out that
similar anomalous (resonant) switching has been observed both experimentally
\cite{Ustinov_03} and theoretically \cite{Jensen_prb} for single-fluxon
behavior in long annular Josephson junctions in an external magnetic field.

\section{Acknowledgment}
This work was supported in part by the Computational Nanoscience Group,
Motorola, Inc, and in part by INFN under the project SQC (Superconducting
Quantum Computing). NGJ acknowledges generous hospitality during several
visits to Department of Physics, University of Rome "Tor Vergata".

\printindex

\begin{thebibliography}{99.}
%
%
%
\bibitem{Barone82}  A.~Barone and G.~Patern\'o, {\it Physics and
Applications of the Josephson Effect} (Wiley, New York, 1982);
T. Van Duzer and C. W. Turner, {\it Principles of
Superconductive Devices and Circuits} , 2nd ed.~(Prentice-Hall, New York,
1998)

\bibitem{Kramers40}  H.~A.~Kramers, Physica {\bf 7}, 284 (1940).

\bibitem{Kramers_stuff}  T. A. Fulton and L. N. Dunkelberger, Phys. Rev. B 
{\bf 9}, 4760 (1974).

\bibitem{Martinis1}  J. M. Martinis, M. H. Devoret, and J. Clarke, Phys.
Rev. Lett. {\bf 55}, 1543 (1985).

\bibitem{Martinis2}  M. H. Devoret, J. M. Martinis, and J. Clarke, Phys.
Rev. Lett. {\bf 55}, 1908 (1985).

\bibitem{Ruggiero99}  B. Ruggiero, M. G. Castellano, G. Torrioli, C.
Cosmelli, F. Chiarello, V. G. Palmieri, C. Granata, and P. Silvestrini,
Phys.~Rev.~{\bf B59}, 177 (1999).

\bibitem{Martinis3}  J. M. Martinis, S. Nam, and J. Aumentado, Phys. Rev.
Lett. {\bf 89}, 117901 (2002).

\bibitem{Han02}  Y. Lu, S. Han, Xi Chu, S. Chu, Z. Wang, Science {\bf 296},
889 (2002).

\bibitem{Berkley03}  A. J. Berkley, H. Xu, M. A. Gubrud, R. C. Ramos, J. R.
Anderson, C. Lobb, and F. C. Wellstood, Physical Review {\bf B68}, 060502
(2003).

\bibitem{Ustinov03}  A. Wallraff, T. Duty, A. Lukashenko, and A. V. Ustinov,
Phys. Rev. Lett. {\bf 90}, 037003 (2003).

\bibitem{Silvestrini97}  P. Silvestrini, V. G. Palmieri, B. Ruggiero, and M.
Russo, Phys. Rev. Lett. {\bf 79}, 3046 (1997).

\bibitem{Jensen_04} N.~Gr{\o}nbech-Jensen, M.~G.~Castellano, F.~Chiarello,
M.~Cirillo, C.~Cosmelli, L.~V.~Filippenko, R.~Russo, and G.~Torrioli,
Phys.~Rev.~Lett.~{\bf 93}, 107002 (2004).

\bibitem{Parisi88}  See, e.g., G.~Parisi, {\it Statistical Field Theory}
(Addison-Wesley, 1988).

%
\bibitem{fabrication}  S. Morohashi and S. Hasuo, J. Appl. Phys. {\bf 61},
4835 (1987).

\bibitem{Affleck}  I. Affleck, Phys. Rev. Lett. {\bf 46}, 388 (1981).

\bibitem{Ustinov_03} A.~Wallraff, A.~Lukashenko, J.~Lisenfeld, A.~Kemp,
M.~V.~Fistul, Y.~Koval, and A.~V.~Ustinov, Nature {\bf 425}, 155 (2003).

\bibitem{Jensen_prb} N.~Gr{\o}nbech-Jensen and M.~Cirillo,
Physical Review B (In Press, 2004) -- cond-mat/0404721.

%
%
%
%
\end{thebibliography}
\end{document}